\documentclass[authoryear,final,5p,showpacs,times,twocolumn]{elsarticle}
\usepackage{amssymb}
\usepackage{amsthm}
\usepackage{natbib}
\usepackage{psfig}
\journal {New Astronomy}
\begin{document}
\begin{frontmatter}
\title{Automated star-galaxy segregation using spectral and integrated band data for 
TAUVEX/ASTROSAT satellite data pipeline
\author[rvt,rvt2]{Archana Bora\corref{cor1}}
\cortext[cor1]{Corresponding author. Present Address: Department of Physics, Gauhati University, Guwahati 781 014}
\ead{archana@iucaa.ernet.in; abora.80@gmail.com}
\author[rvt]{Ranjan Gupta}
\author[rvt1]{Harinder P. Singh}
\author[rvt2]{K. Duorah}
\address[rvt]{IUCAA, Post Bag 4, Ganeshkhind, Pune-411007, India\fnref{label3}}
\address[rvt1]{Department of Physics and Astrophysics, University of Delhi, Delhi 110007, India\fnref{label3}}
\address[rvt2]{Department of Physics, Gauhati University, Guwahati 781014, India\fnref{label3}}
\fntext[label3]{}
 }

\begin{abstract}

We employ an Artificial Neural Network (ANN) based technique
to develop a pipeline for automated segregation of stars from the galaxies to be observed by 
Tel-Aviv University Ultra-Violet Experiment (TAUVEX). We use
synthetic spectra of stars from UVBLUE library and selected International Ultraviolet
Explorer (IUE) low resolution spectra for galaxies
in the ultraviolet (UV) region from 1250 to 3220\AA~ as the training set and
IUE low-resolution spectra for both the stars and the galaxies as the test set.
All the data sets have been pre-processed
to get band integrated fluxes so as to mimic the observations of the TAUVEX UV imager. 
We also perform the ANN based segregation scheme using the full length spectral features
(which will also be useful for the ASTROSAT mission).
Our results suggest that, in the case of the non-availability of full spectral features, 
the limited band integrated features can be used to segregate the two classes of objects;
although the band data classification is less accurate than the full spectral data
classification.
\end{abstract}

\end{frontmatter}

\section{Introduction}
The TAUVEX (Tel-Aviv University Ultra-Violet Experiment) is a collaborative 
UV imaging experiment between Indian Institute of Astrophysics (IIA, Bangalore,
India) and Tel-Aviv University, Israel. 
TAUVEX will have three UV imaging telescopes of 20-cm area which will
obtain UV images of the sky in the spectral region of 1250\AA~ to 3220\AA~ with
5 different band filters. 
Each telescope has a field of view (FOV) of about 54'
and a spatial resolution of about 6-10", depending on the wavelength.
It will be placed into a geostationary orbit as part
of Indian Space Research Organization's (ISRO's) GSAT-4 mission.

In its life time the satellite is expected to
collect data of about $10^{6}$ celestial sources (Brosch 1998)
of both point sources (stars, QSO's etc) as well as extended sources like nebulae, 
galaxies, clusters etc.
With such a huge data set there is a call for automatizing the segregation
of the point sources from other non-point-like sources. 
Different machine learning algorithms governed by some 'learning rule' serves this purpose
and they are now routinely used in astronomy for different classification problems.
There are three major learning paradigms for machine learning algorithms: 
supervised learning, unsupervised learning and reinforcement learning, each 
corresponding to a particular learning task. Some examples of such machine learning 
algorithms are artificial neural network (ANN), support vector machines, difference boosting neural network, self-organized maps (SOM) etc.

Odewahn et al. (1992) pioneered the use of ANN based scheme for automated segregation 
of stars and galaxies based on point-spread function (PSF) fitting.
M\"a{h}\"o{n}en et al. (1995) used SOM based neural network using the
CCD images directly. M\"a{h}\"o{n}en et al. (2000) also introduced another method based on 
fuzzy set reasoning.
Philip et al. (2002) used the difference boosting
neural network for the star-galaxy classification of NOAO Deep
Wide Field Survey (NDWFS).
Qin et al. (2003) also
demonstrated the use of spectra for the 
same purpose using RBF neural network, in the wavelength range of 3800-7420\AA~.
In this paper, we propose the use of ANN with integrated flux measurements 
in different bands,
in case of non-availability of full spectra, to separate objects (stars and galaxies)
with different spectral energy distributions. The wavelength of concern in the present 
work is the UV range (1250-3200\AA). 
We also classify the two types of objects using corresponding full spectral informations and 
present a 
comparison between the two schemes (i.e, using band data and full spectral data).

It is to be noted here that the integrated flux approach has already been used to 
classify stellar objects into different spectral types and to estimate the colour 
excess for the hot stars (Bora et al. 2008).
However, while doing the classification in the above mentioned work, it was assumed that the separation 
of the stellar objects from other celestial objects have been done apriori by some other
method. Thus with the incorporation of the proposed scheme
both the tasks of star-galaxy segregation and star classification can be performed with 
the band integrated flux to be available from the TAUVEX satellite.

\section{Network Architecture}

In this section, we very briefly review the basics of ANN and describe
the network structure used in the present work.
The neural network considered here is the supervised learning network,
with a back-propagation algorithm (Gulati et al. 1994, 1997a,b; Singh et al. 1998).
The network consists of three layers, i.e. (i) input layer, (ii) hidden layer
and (iii) output layer.
Patterns are presented to the network for learning at the input layers which
are subsequently communicated to the hidden layer.
The hidden layer interconnects the input and the output layers and can have several
nodes with a particular transfer function.
The actual processing is done in the hidden layer via weighted connections
and the outputs are rendered at the output layers (Bailer-Jones, Gupta \& Singh 2002),
We have used two hidden layers of 64 nodes each with a Sigmoid function
as the transfer function. The scheme requires a training session
where the ANN output and the desired output get compared
after each iteration and the connection weights get updated till the
desired minimum error threshold is reached. At this stage, the network
training is complete and the connection weights are considered
frozen. The next stage is the testing session in which the test patterns
are fed to the network and output is the classification of the objects
as star or galaxy. 

\section {Data for training and testing}

In the following, we describe the data sources for the stellar and galaxy spectra, and the generation
of the train and test test sets for the network.

\subsection {The sources for stellar and galaxy spectra}
\textit{(i) The stellar data set:} We have used the UVBLUE fluxes (Rodriguez-Merino et al. 2005) for generating the 
training sets for stellar spectra with solar type stars with $[M/H]$=$0$ (http://www.bo.astro.it/$\sim$eps/uvblue/uvblue.html).
The references (Allen (2000), Erika B\"e{o}hm-Vitense (1981), Johnson (1966), Ridgway et al. (1980), Alonso, Arribas \& Martinez-Roger (1999) and Bertone et al. (2004))
provide the 
necessary information for matching the parameter space of UVBLUE to spectral-types.
The UVBLUE library source provides the sets of theoretical 
fluxes (based on Kurucz model atmospheres) in the UV region.
 
The test spectra were taken from the IUE low resolution spectra:
reference atlas, normal stars, ESA SP-1052 by Heck et al. (1984)
which contains 229 low-dispersion flux calibrated spectra of O to K
spectral type at a resolution of 6\AA~ obtained by the IUE satellite.

\textit{(ii) The galaxy data set:} For galaxy spectra we have used the UV-optical spectra of 99 Nearby Quiescent and Active Galaxies available online on http://www.stsci. edu/ftp/catalogs/nearby\_gal/sed.html (Storchi-Bergmann et al.).
The spectra covers the wavelength range 1200-3200 \AA~ with a resolution of 5-8\AA~.

Although an extended grid of synthetic galaxy spectra are available in the wavelength
range 2500 to 10500 \AA~ with zero-redshift (Fioc et al. 1997), there is no 
such grid in the range of our interest (1200-3200\AA~). So, we choose a subset of the galaxies from the above 
mentioned database as the train set, keeping the rest
for the test set. The selection of the galaxy subset for the training session
is based on the following general observations
related to the galaxy spectra:

The spectra of galaxies of different morphological types 
reveal that the Elliptical and the S0 galaxies essentially have no
star-formation activity and the spectra of these two types look quite
similar. By contrast, the Sc spirals and the Irr galaxies have a spectrum dominated
by emission lines. On the other hand the Sa and the Sb galaxies form
a kind of transition between these early-type galaxies and
Sc galaxies (Schneider 2006). 
Depending on these spectral distributions 
we prepare a list of 16 galaxies spanning over all the above 
mentioned types as the training sample. The remaining 69 galaxies are then used 
as the test set.
All these galaxies considered
in the train set and in the test are at the low red-shift regions.

It is to be mentioned that the TAUVEX detectors being virtually noiseless and also
with little or no stray light in parts of the orbit, its detection will be
limited only by photon statistics
(Brosch 1998; Safonova et al. 1998) and so the effect of noise on the data is not taken 
into account.

\begin{figure}
\centering
\includegraphics[width=6.5cm]{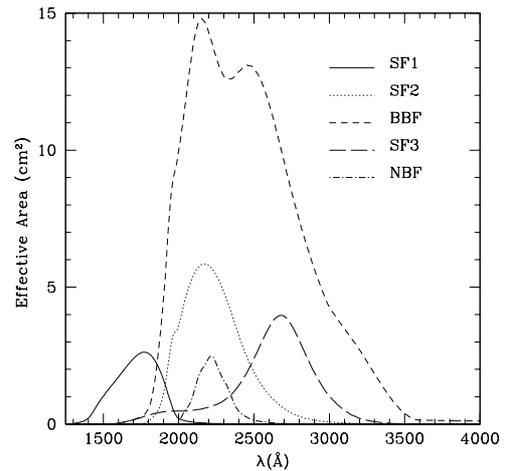}
\caption{Filter response of the five filters of TAUVEX}
\end{figure}

\subsection{Generation of the Train and the Test sets:}

While making the train and the test sets, one has to ensure that the number of spectral 
fluxes at the respective wavelengths and the starting, ending wavelengths are 
identical. Also the spectral resolution of the two sets needs to be the same.
This has been achieved by first trimming the spectra
in the range of 1250-3200\AA~ at 40 data bins and then 
convolving the IUE star/galaxy spectra with appropriate Gaussian functions
to bring down their resolution to 50\AA.
The resolution of UVBLUE spectral types have also been degraded to a resolution of 50\AA~
using the relevant codes provided on the UVBLUE library web site (http://www.bo.astro.it/~eps/uvblue/go.html). 
 
The details of the procedure adopted for generating the band integrated train and test sets is
described below with reference to the TAUVEX filter response.
For the TAUVEX mission, the observations will be available from 1250\AA~ to 3220\AA~
spectral region using filters, namely BBF, SF1, SF2, SF3 and NBF3, in five UV bands.
Fig.$1$ shows the total response of each of the TAUVEX filters  
in units of Effective Area $\rm cm^{2}$ and their approximate characteristics 
are summarized in Table 1.

\begin{table}
\caption{TAUVEX filters specifications}
\begin{center}
\begin{tabular}{lclclcl}
\hline
Filter & Wavelength &    Width & Normalized\\
 & \AA~ & \AA~  &  transmission\\
\hline
BBF &   2300   &       1000 & 80\% \\
SF1 &   1750   &       400  & 20\% \\
SF2 &   2200   &       400  & 45\%\\
SF3 &   2600   &       500  & 40\%\\
NBF3&   2200   &       200  & 30\%\\
\hline
\end{tabular}
\label{tab1}
\end{center}
\end {table}

The spectra are degraded to same resolution and rebinned to a common
spectral range and then the fluxes are processed via a common flux 
integration programme provided at the TAUVEX tool site to form
two sets of band data (each having five fluxes corresponding to the four TAUVEX bands).

We have also obtained two sets of fluxes (with 50\AA~ resolution
and 40 data bins covering the spectral region of 1250-3220\AA) aimed at 
preparing the ANN tools for another Indian scientific
mission satellite ASTROSAT (http://www.rri.res.in/astrosat/)
which will have gratings to provide slitless spectra for spatially
resolved stars. It will also prepare us for the future GAIA mission
(http://gaia.esa.int/science-e/www/area/index.cfm?fareaid=26).
For this the spectra are normalized to unity with respect to the maximum flux
in each spectrum before sending to the network.
Fig. 2 and Fig. 3, shows the block diagram for creating the 
UVBLUE train sets and the IUE test sets for the star while Fig. 4 shows the 
flow chart for creating the IUE train and the test sets for the galaxy.

\begin{figure}
\centering
\includegraphics[width=8.5cm]{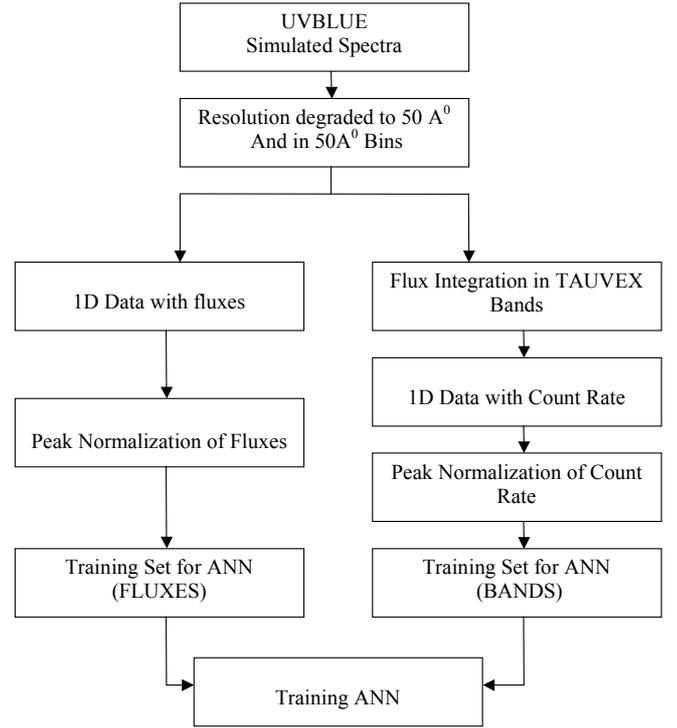}
\caption{A block diagram showing the flow chart for creating the ANN
train set for star with UVBLUE simulated sources.}
\end{figure}

\begin{figure}
\centering
\includegraphics[width=8.5cm]{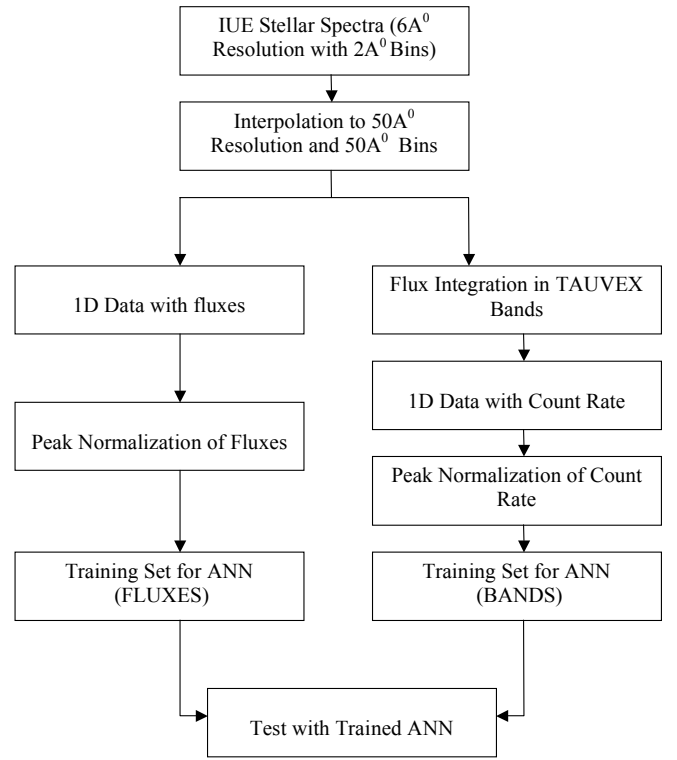}
\caption{A block diagram showing the flow chart for creating the ANN
test set for stellar sources}
\end{figure}

\begin{figure}
\centering
\includegraphics[width=8.5cm]{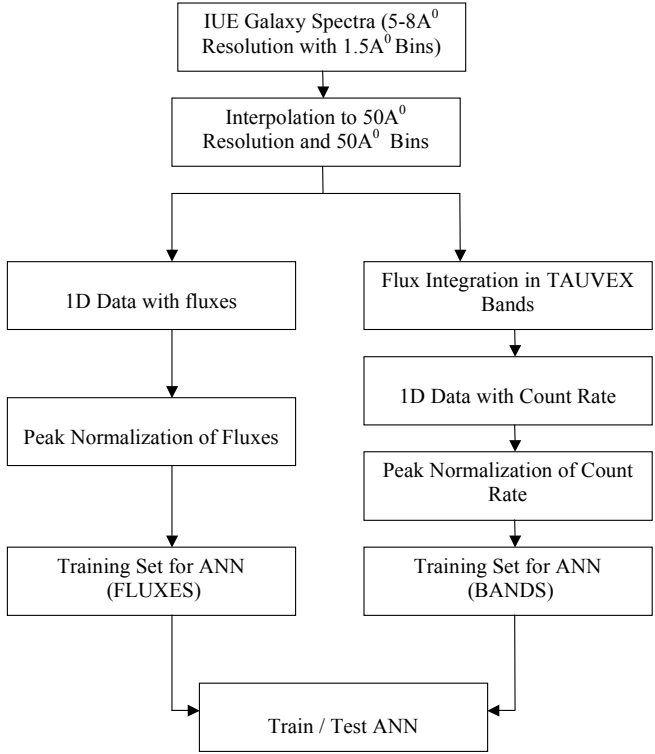}
\caption{A block diagram showing the flow chart for creating the ANN
train/test set for galaxy}
\end{figure}

In the current work, we have neglected the effect of red-shift on the galaxy
spectra and this will be incorporated in a future upgrade of the software
which is being developed.

\section[]{Results of the classification}

\subsection{Band Data}

Separating the stellar objects from the galaxies using the five TAUVEX data points only
as the classification features, is a challenging job. 
However, the use of all the available informations in the five
filters enable us to classify 170 (128 stars and 42 galaxies)
objects correctly out of a sample of 297 stars plus galaxies,
yielding the success rate of classification as 57\%.
The scattered 3D plot of the classification \% in the sub-space
is shown in Fig. 5. In these three dimensional
scatter plots, the two axes in the horizontal plane denote the catalog
and ANN classes, and the vertical axis in the plots
gives the number of stars present for a particular class.
In the plot stars are labeled as 1000 while galaxies are labeled as 2000.

\begin{figure}
\centering
\includegraphics[width=8.5cm]{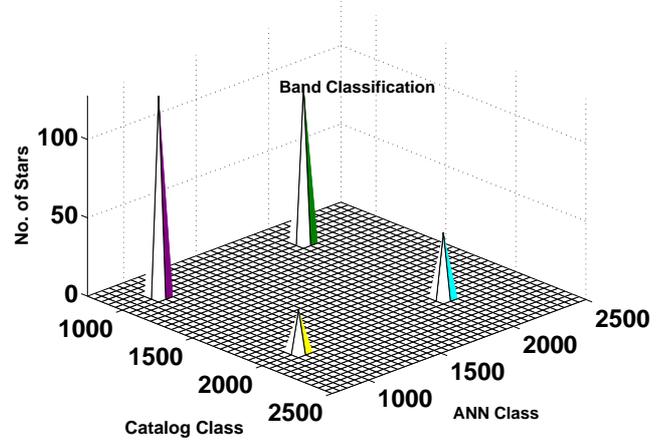}
\caption{Scatter plot of classification of 229 IUE stars and 69 galaxies using Band data;
(class-code: 1000-stars; 2000-galaxy)}
\end{figure}

\subsection{Flux data}

The classification results with the full spectral
features excel the result with the band integrated features, which is to be expected as
the full spectra will always have more information as compared to the
band data.
The network classifies 226 objects correctly, 
(171 being stars and 55 being galaxies) out of a sample of 297 stars and galaxies,
yielding the success rate of classification as 76\%.
The scattered 3D plot of the classification is shown in Fig. 6. 
The result of both the band classification and
the flux classification is summarised in Table 2.

\begin{figure}
\centering
\includegraphics[width=8.5cm]{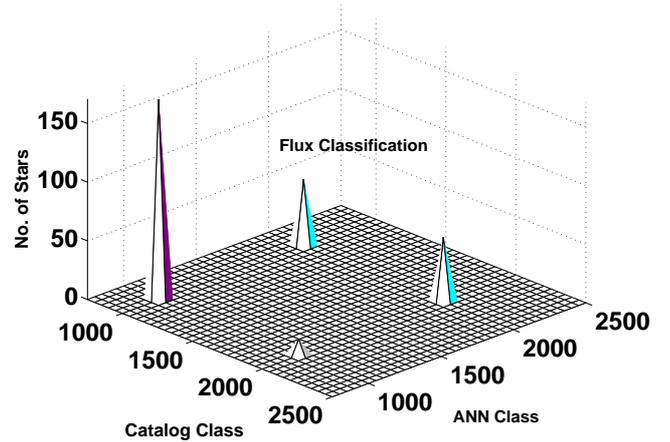}
\caption{Scatter plot of classification of 229 IUE stars and 69 galaxies using Flux data;
(class-code: 1000-stars; 2000-galaxy)}
\end{figure}

\section[]{Conclusions}

Generally stars are separated from the galaxies by PSF fitting
and this has been extensively done in the optical region of the electro magnetic spectrum. 
The extension of applicability of the neural network based scheme to the UV
region has been less prevalent mainly because of non-availability of
abundant data in this region. The present work demonstrates
that the ANN can be successfully employed to
separate stars from galaxies in the UV region too.
We have shown that the tool developed by us can successfully
classify the two classes of objects using 
both their full spectral information as well as their photometric data.
The success rate is 76\% when using the Flux data.
Automatizing the
separation of star and galaxy using only few photometric data 
is indeed a challenging job
to perform.
The result of our analysis shows that with our ANN tool 
this can be achieved with a success rate of 57\%.
Thus, even with the limitation of data from just five photometric
bands, the ANN can be used to classify the point sources from the extended sources
like galaxies/AGNs etc.
There is a good chance for improving the classification results with larger
data sets to be available from the upcoming satellite missions TAUVEX/ASTROSAT/GAIA etc.

\begin{table} 
\caption{Classification and Results.}
\begin{center}
\begin{tabular}{lll}
\hline
Classification Scheme &\hspace{0.02in} Results \\
\hline
Band & Star:\hspace {0.14in} 128/228 \\
 & Galaxy: \hspace{0.12in}42/69 \\
 \hline
Flux & Star:\hspace{0.14in} 171/228 \\
 & Galaxy:\hspace{0.12in} 55/69 \\
 \hline
\end{tabular}
\label{tab1}
\end{center}
\end{table}

\section*{Acknowledgments}
This work is supported by a research grant from Indian Space Research Organization under its 
RESPOND scheme. AB would like to thank Prof. S. Ravindranath for suggesting useful references
that helps to understand the galaxy spectra.


\begin{thebibliography}{}
\bibitem{Allen} Allen 2000, Astrophysical Quantities, 4th edn, Springer-Verlag, New York
\bibitem{Alonso} Alonso, A., Arribas, S., Martinez-Roger, C., 1999, A\&AS, 140, 261.
\bibitem{Bailer01} Bailer-Jones, C. A. L. 2002, in Automated Data Analysis in Astronomy, ed. R. Gupta, H. P. Singh, \& C. A. L. Bailer-Jones (New Delhi: Narosa), 83.
\bibitem{Bailer02} Bailer-Jones, C. A. L., Gupta, R., \& Singh,H. P. 2002, in Automated Data Analysis in Astronomy, ed. R. Gupta, H. P. Singh, \& C. A. L. Bailer-Jones (New Delhi: Narosa), 51.
\bibitem{Bertone} Bertone E., Buzzoni A., Rodriguez-Merino L. H., Chavez M., 2004, AJ, 128, 829.
\bibitem{Bora} Bora, A., Gupta, R., Singh, H. P., Murthy, J., Mohan, R., Duorah, K., 2008. MNRAS, {\bf384}, 449.
\bibitem{Brosch} Brosch, N., 1998. Physica Scripta-Supplement T, {\bf77}, 16.
\bibitem{Erika} Erika B\"e{o}hm-Vitense, 1981. Ann. Rev. Astron.  Astrophys., 295, 318.
\bibitem{Fioc} Fioc M., Rocca-Volmerange B., 1997. A\&A, {\bf326},950.
\bibitem{Gulati} Gulati, R.K., Gupta, R., Gothoskar, P., Khobragade, S., 1994. ApJ, {\bf426}, 340 (1994).
\bibitem{Gulati2a} Gulati, R. K., Gupta, R., Rao, N.K., 1997a. A\&A, {\bf322}, 933.
\bibitem{Gulati2b} Gulati, R.K., Gupta, R., Singh, H.P., 1997b. PASP, {\bf109}, 843.
\bibitem{Heck} Heck A., Egret D., Jaschek M., Jaschek C., 1984. ESA SP-1052, IUE lowdispertion spectra reference atlas. Part 1. Normal Stars, ESA SP-1052.
\bibitem{Johnson} Johnson, H. L., 1966. Ann. Rev. Astron.  Astrophys., 4, 193.
\bibitem{Mahonen1} M\"a{h}\"onen, P., Hakala, P. , 1995. ApJ {\bf 452}, L77.
\bibitem{Mahonen2} M\"a{h}\"onen, P., T. Frantti, 2000. ApJ {\bf 541}, 261. 
\bibitem{Odewahn} Odewahn,S. C., Stockwell, R. L. Pennington, R. L., Humphreys, R., M., Zumach, W. A., 1992. The Astronomical Journal, {\bf103}, 318.
\bibitem{Philip} Philip, N. S., Wadadekar, Y., Kembhavi, A., Joseph, K. B., 2002. A\&A, {\bf385}, 1119.
\bibitem{Schneider} P. Schneider, \textsl{Extragalactic Astronomy and Cosmology: An Introduction}, (Springer-Verlag, Berlin, Heidelberg, 2006).
\bibitem{Qin} Qin, D.M., Guo, P., Hu, Z.U., \& Zhao, Y.H., 2003. Chin. J. Astron. Astrophys, {\bf3}, 277.
\bibitem{Ridgway} Ridgway, S. T., Joyce, R. R., White, N. M. and Wing, R. F., 1980, ApJ, 235, 126
\bibitem{Rodriguez} Rodriguez-Merino, L. H., Chavez, M.,Bertone, E., Buzzoni, A. 2005, ApJ, 626, 411 (UVBLUE)
\bibitem{Safonova} Safonova, M, Sivaram, C., \& Murthy, J., 2008. Astrophys Space Sci, {\bf318}, 1.
\bibitem{Singh} Singh, H.P., Gulati, R.K., Gupta, R., 1998. MNRAS, {\bf295}, 312.
\bibitem {Bergmann} Storchi-Bergmann, T., Calzetti, D.,  Kinney, A. L., \textsl{A database of UV-optical spectra of nearby quiescent and active galaxies}, (http://www.stsci.edu/ftp/catalogs/nearby\_gal/sed.html).
\end{thebibliography}
\end{document}